%% file: main.tex
\documentclass[aps,prl,twocolumn,superscriptaddress]{revtex4-1}
\usepackage{natbib}
\usepackage{graphicx}
\usepackage{wrapfig}
\usepackage{amsmath}
\usepackage{latexsym}
\usepackage{amssymb}

\begin{document}

\title{Observation of Transverse Spin-Dependent Azimuthal Correlations of Charged Pion Pairs in $p^\uparrow+p$ at $\sqrt{s}=200~GeV$}


\input{authors}


\date{\today}

\begin{abstract}
We report the observation of transverse polarization-dependent azimuthal correlations in charged pion pair production with the STAR experiment in $p^\uparrow+p$ collisions at RHIC.
These correlations directly probe quark transversity distributions.
We measure signals in excess of five standard deviations at high transverse momenta, at high pseudorapidities $\eta>0.5$, and for pair masses around the mass of the $\rho$-meson.
This is the first direct transversity measurement in {\it p+p} collisions. Comparing the results to data from lepton-nucleon scattering will test the universality of these spin-dependent quantities.


\end{abstract}


\maketitle


The non-perturbative structure of the nucleon can be described in terms of parton distribution functions (PDFs), equivalent to number densities of quarks and gluons in a fast moving nucleon.
Transversity, $h^q_1(x)$, is the least well known of the PDFs.
It represents the transverse quark polarization in transversely polarized nucleons for quark flavor $q$ and momentum fraction $x$.
Due to its chiral odd nature, transversity vanishes for gluons in the nucleon ($s_z=\frac{1}{2} \hbar$) and is primarily a property of the valence quarks~\cite{Barone:2001sp}.
An experimental measurement of the nucleon tensor charge $\delta q=\int_0^1 dx (h^{q}_{1}(x)-h^{\bar{q}}_{1}(x)) $ will directly test our theory of quantum chromodynamics (QCD) when compared to calculations on the lattice or model calculations ~\cite{Anselmino:2013vqa,Gamberg:2001qc,Bacchetta:2012ty,Cloet:2007em,Wakamatsu:2007nc,Gockeler:2005cj,He:1996wy,Schweitzer:2001sr,Pasquini:2005dk,Bacchetta:2008af}.
$h_1$ becomes accessible in physics observables when it is coupled with an additional chiral-odd partner, e.g. a transverse spin-dependent fragmentation process.
This second part has to be measured independently in order to extract $h_1$.
Our current knowledge of $h_1$~\cite{Anselmino:2013vqa,Bacchetta:2012ty} is based on fixed-target semi-inclusive deep inelastic lepton-nucleon scattering (SIDIS)~\cite{Airapetian:2004tw,Airapetian:2008sk,Adolph:2012nw,Alekseev:2010rw,Alekseev:2008aa} in combination with data from electron-positron annihilation ~\cite{Seidl:2008xc,Vossen:2011fk}.
Proton-proton collisions allow us to reach into the dominant valence quark region, but the framework of perturbative QCD introduces complications when the intrinsic transverse momentum from the hadronization process has to be considered~\cite{Rogers:2010dm}.
It has been shown that di-hadron correlations in the final state persist when integrated over intrinsic transverse momenta.
This so-called Interference Fragmentation Function (IFF), $H_1^{\sphericalangle}$, can therefore be described collinearly~\cite{PhysRevLett.80.1166}. Therefore the contributions to the cross section can be factorized~\cite{Boer:2003ya}
and the IFF should be universal between electron-positron annihilation, SIDIS, and proton-proton scattering.

We present measurements of charged pion correlations from the STAR experiment at the Relativistic Heavy Ion Collider (RHIC) at a center-of-mass energy $\sqrt{s}=200$~GeV.
The data, the first measurement of transversity in polarized proton collisions, show non-zero $h^q_1(x)$ at $0.15<x<0.30$.
In this range, transversity is not well constrained by previous SIDIS measurements and our result will be particularly 
important to restrict the $d$-quark transversity which is charge suppressed in lepton-proton scattering.

RHIC, located at Brookhaven National Laboratory, collides bunched beams of heavy ions as well as polarized protons.
The stable beam polarization orientation is transverse to the collider plane and the polarization direction alternates between subsequent bunches or pairs thereof (polarization up $\uparrow$ or down $\downarrow$).
The bunch polarization pattern is changed from fill to fill in order to reduce systematic effects.
While typically both beams are polarized, a single-spin measurement is achieved by summing over the bunches in one beam, effectively reducing its polarization to near zero.
The polarization of each beam is measured by polarimeters using the elastic scattering of protons on very thin carbon targets, several times during a RHIC fill.
The polarimeter are calibrated using a polarized hydrogen gas jet target~\cite{polarimeter}.
We report results from the RHIC run in 2006 with an integrated luminosity of 1.8~pb$^{-1}$ and an average beam polarization of about 60\%.

The STAR experiment is located at one of the collision points in RHIC.
This analysis is based on data in the central pseudorapidity range $-1<\eta<1$.
Data are collected by the Time Projection Chamber (TPC) providing tracking and charged pion identification~\cite{Ackermann:2002ad} and by the Barrel Electromagnetic Calorimeter (BEMC), a lead scintillator sampling calorimeter~\cite{Beddo:2002zx}.
Data from a pair of scintillator-based beam-beam counters (BBC) at forward rapidities $3.3<|\eta|<5.0$ in combination with the BEMC provides a trigger for hard QCD events~\cite{Da:2011da}.
The trigger requires a coincidence between the BBCs and either a minimum transverse energy, $E_T>5$~GeV in a single BEMC tower or one of several jet patch triggers in $\Delta\phi\times\Delta\eta=1.0\times 1.0$ ($E_T>$ 4.0 or 7.8 GeV).

Charged pion pairs are selected by requiring tracks that originate
within $\pm 60$~cm in the longitudinal direction and 1~cm in the transverse direction from the nominal interaction vertex and that are required to point into the central region.
Tracks are required to have a minimum transverse momentum $p_T$ of 1.5 GeV/$c$. Using $dE/dx$ measurements in the TPC to select pions, a purity of the single pion sample of greater than 95\% over the whole kinematic range is achieved.
All pion pairs in an event are considered where the pions are close enough in ($\eta,\phi$) space to originate from the fragmentation
of the same parton. The default value of this opening angle cut is $\sqrt{(\eta_{\pi_1}-\eta_{\pi_2})^2+(\phi_{\pi_1}-\phi_{\pi_2})^2}<0.3$. 
Pion pairs produced in the weak decay of the $K^0$ meson are not expected to contribute to the asymmetry, therefore the corresponding mass range (497.6 $\pm$ 10 MeV) was excluded from the analysis.







\begin{figure}[h]
\includegraphics[width=0.5\textwidth]{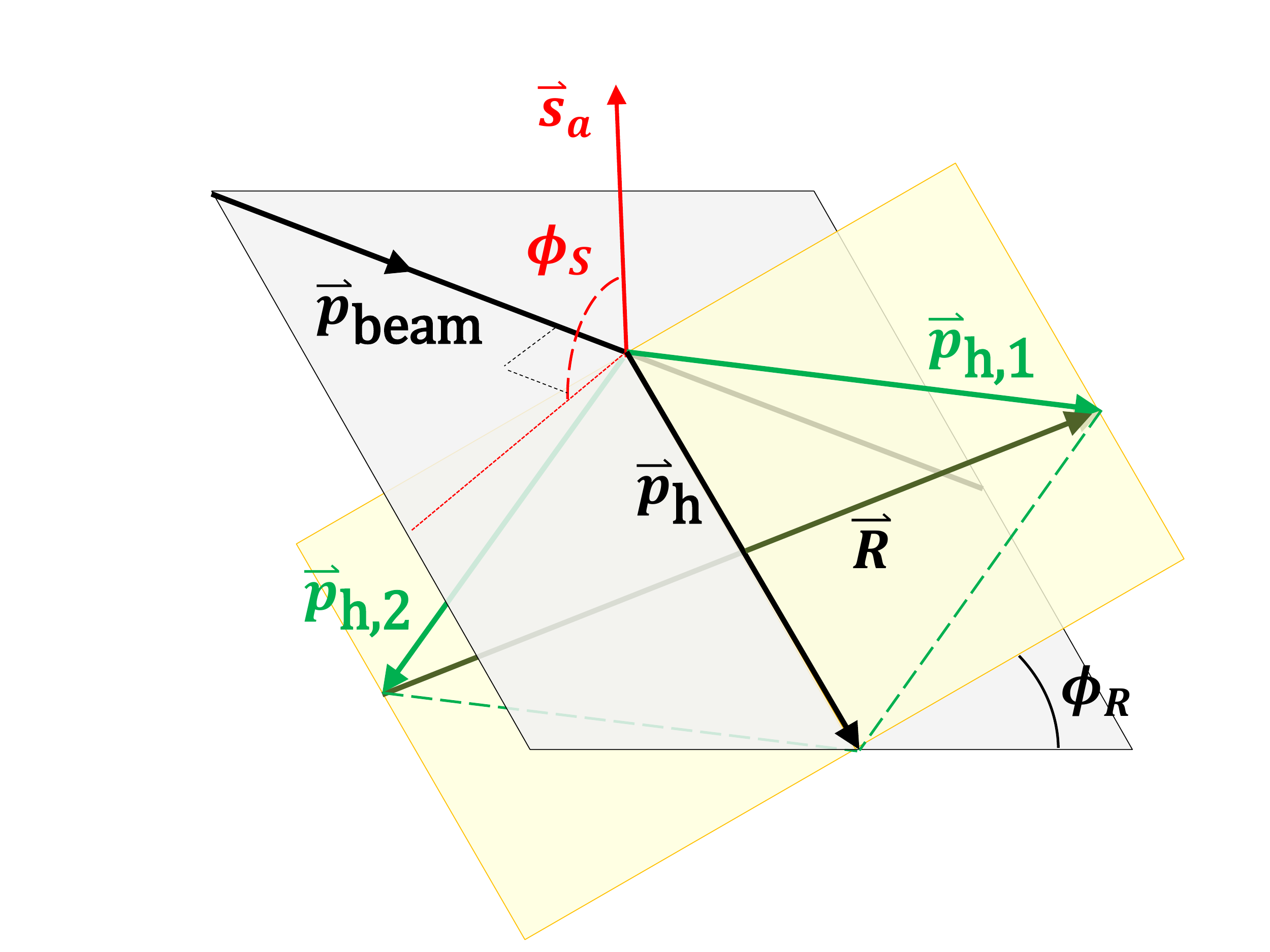}
\caption{Azimuthal angle defintions in the dihadron system.
$\vec{s}_{a}$ is the direction of the spin of the polarized proton, $\vec{p}_{h,\{1,2\}}$ are the
momenta of the positive and negative pion, respectively and $\phi_R$ is
the angle between the production and dihadron plane.
\label{fig:angles}}
\end{figure}

The transversely polarized cross-section of hadron pairs in $p^\uparrow+p$ collisions can be written similar to~\cite{Bacchetta:2004it}:
\begin{equation}
d\sigma_{UT}\propto \sin(\phi_{RS})
\int dx_a dx_b f_1(x_a) h_1(x_b) \frac{d\Delta\hat{\sigma}}{d\hat{t}} H_{1,q}^\sphericalangle(z,M).
\label{eq:xsect}
\end{equation}
Here, $\hat{\sigma}$ is the polarized scattering cross section of partons $a$ and $b$ with four momentum transfer $\hat{t}$. The unpolarized parton distribution is $f_1(x)$.
The fragmentation function $H_{1,q}^\sphericalangle$ is a function of $z$, the fractional energy with respect to the fragmenting quark  carried by the hadron pair and its invariant mass, $M$.
The angle $\phi_{RS}=\phi_R - \phi_S$ is derived according to Fig.~\ref{fig:angles} from the angle between the polarization vector and the production plane, $\phi_S$ and the angle between the two hadron plane and the production plane, $\phi_R$.
The production plane is spanned by the incident proton momentum, $\vec{p}_{beam}$, and the sum of the two hadron momenta, $\vec{p}_h=\vec{p}_{h,1}+\vec{p}_{h,2}$.
The difference of the momenta $\vec{R}=\vec{p}_{h,1}-\vec{p}_{h,2}$ lies in the hadron plane. 
The convolution of $h_1(x)$ and $H_{1,q}^{\sphericalangle}$ will introduce an asymmetry, modulated by $\sin(\phi_{RS})$.
The effect will inherit the dependence on the partonic variable $x$ from $h_1(x)$ and the final state variables $M$ and $z$.


An experimental observable directly proportional to the differential cross-section is constructed for each RHIC fill:
\begin{equation}
\frac{N^\uparrow(\phi_{RS})-r\cdot N^\downarrow(\phi_{RS})}{N^\uparrow(\phi_{RS})+r\cdot N^\downarrow(\phi_{RS})} = P_{beam}\cdot A_{UT}\cdot\sin(\phi_{RS}).
\label{eq:simpleForm}
\end{equation}
where $N^{\uparrow/\downarrow}$ is the number of pion pairs meeting the selection criteria for each polarization state, $P_{beam}$ the beam polarization and  $r$ the ratio $L^{\uparrow}/L^{\downarrow}$ between the integrated luminosities of the two polarization states.


The data is binned in 16 equal bins covering $2\pi$ in azimuth. The amplitude $A_{UT}$ of $\sin(\phi_{RS})$ is extracted by a fit to the data.
The description of the functional form is very good, with a reduced $\chi^2$ per degree of freedom of $0.975\pm 0.007$ over all kinematic bins.
We include all pion pairs with opposite charges from an event and define $\vec{p}_{h,1}$ to be the momentum of the positive particle (and $\vec{p}_{h,2}$ the negative particle accordingly).
Note that this charge ordering is essential because it establishes the direction of $\vec{R}$.
A random charge assignment would lead to a vanishing asymmetry since it would randomize the sign of $\phi_{RS}$.


\begin{figure}[h]
\begin{center}
	\includegraphics[width=0.45\textwidth]{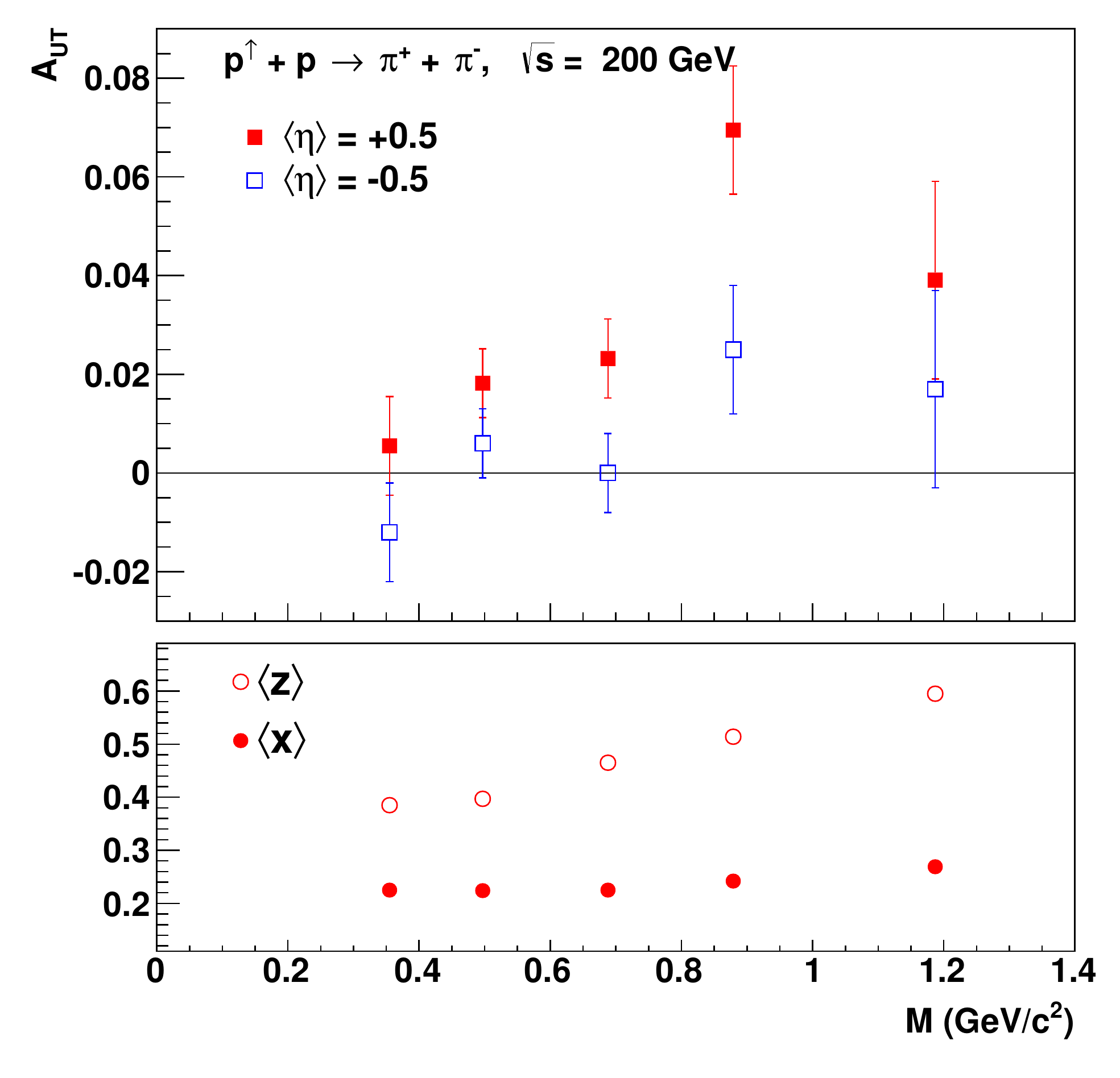}
	\caption{$A_{UT}$ as a function of $M$ (upper panel) and corresponding partonic variables $x$ and $z$ (lower panel).
    A clear enhancement of the signal around the $\rho$  mass region can be observed.\label{fig:fwdBwdAsymM} }. 
\end{center}
\end{figure}

Figure~\ref{fig:fwdBwdAsymM} shows the results for $A_{UT}$ as a function of the invariant mass $M$ of the pion pair, both for forward ($\eta>0$)and backward ($\eta<0$) going particles.
We define the forward direction here along the momentum of the polarized beam.
The results combine independent measurements of the asymmetries for both polarized RHIC beams in the two halves of the STAR detector, which provides internal consistency checks.


We used a Pythia~\cite{Sjostrand:2006za} simulation in conjunction with a  model of the STAR detector response implemented in GEANT~\cite{Brun:1978fy} to determine the partonic scattering processes as well as the partonic variables $x$ and $z$, the fractional momentum of the parent quark carried by the two hadrons.
These are shown in the lower panel of Fig.~\ref{fig:fwdBwdAsymM}.
The simulation agrees reasonably well with the data and detector resolutions are small compared to the presented bin sizes.
The mean $x$ value, $\langle x\rangle$, of the recorded data at midrapidity is around 0.2 and it changes very little over the available invariant mass range.
This is well in the valence region, $x>0.1$, where transversity is expected to be sizable.
On the other hand, $\langle z\rangle$ rises more strongly with the invariant mass.
This is essentially a consequence of the opening angle cut and the required minimum $p_T$ for each hadron.
Naively one expects that the IFF is uniformly rising in $z$, since hadrons at high $z$ carry more of the parent quark spin information.
This is consistent with measurements in $e^+e^-$ annihilation ~\cite{Vossen:2011fk} where sizable values have been observed at similar $z$ and $M$.

In model calculations, the transverse spin dependence of the IFF originates from an interference of amplitudes with different angular momenta~\cite{Bianconi:1999uc}.
In our kinematic region, this will mainly be contributions from vector meson decays in a relative p-wave which interfere with non-resonant background in a relative s-wave.
Therefore, it is expected that the invariant mass dependence will show an enhancement around the mass of the $\rho$ meson~\cite{Bianconi:1999uc}.
Our results confirm these expectations and show a clear signal in the forward direction around the $\rho$ mass.

Backward asymmetries, $\eta<0$, are sensitive to quarks at small $x$.
They are consistent with zero, as is expected since transversity is primarily carried by the valence quarks.

\begin{figure}[h!]
\begin{center}
	\includegraphics[width=0.45\textwidth]{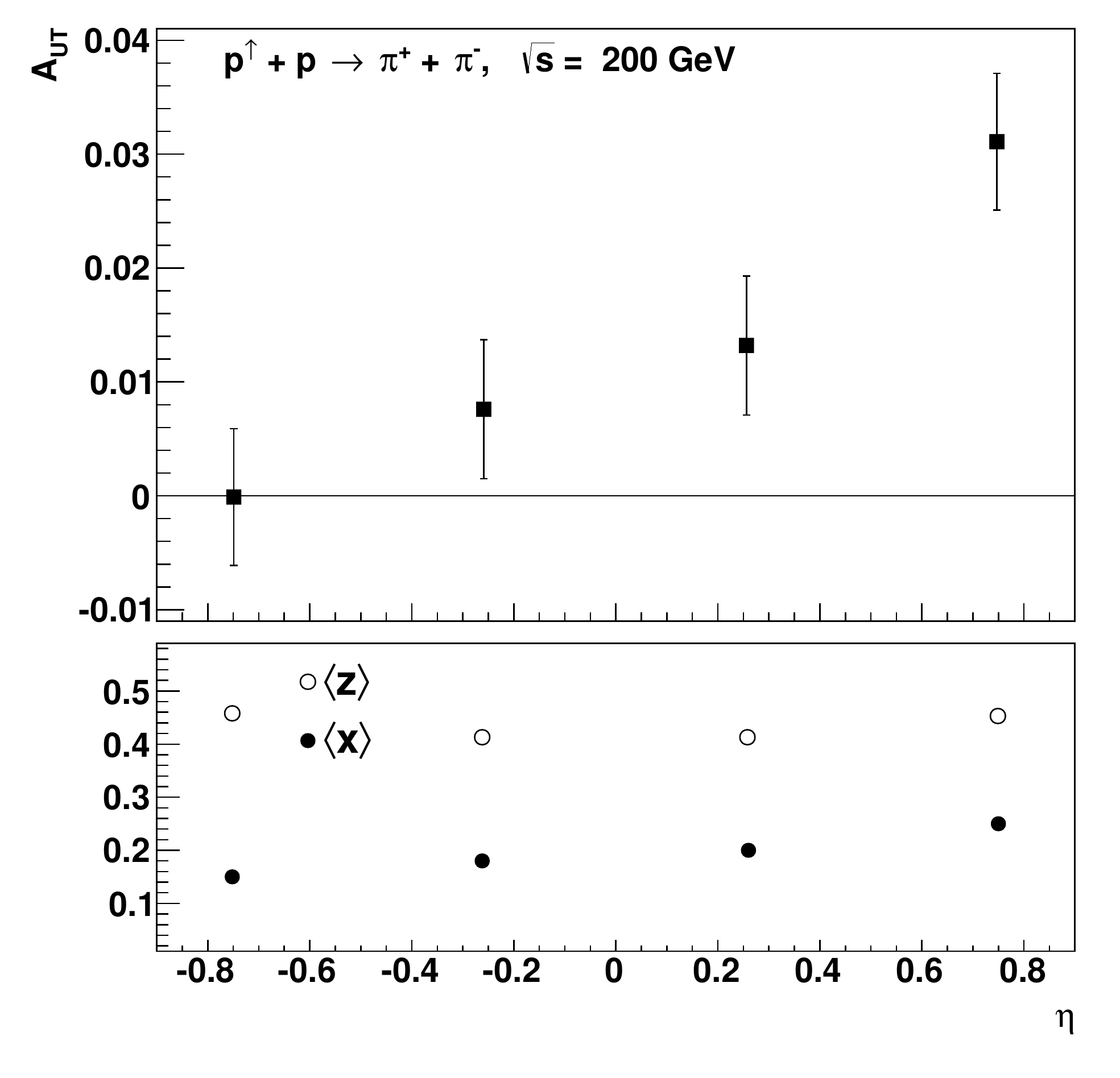}
	\caption{$A_{UT}$ as a function of $\eta$ (upper panel) and corresponding partonic variables $x$ and $z_{\pi^+\pi^-}$ (lower panel). \label{fig:etaAsym}}
\end{center}
\end{figure}


Figure~\ref{fig:etaAsym} shows $A_{UT}$ as a function of $\eta$ in more detail.
The $\langle x\rangle$ of the polarized beam rises approximately linearly with $\eta$ from 0.15 to 0.25 while $\langle z\rangle \approx 0.4$ in the covered acceptance.
The measured asymmetries reflect the $x$-dependence and valence quark nature of transversity and rise monotonically with $\eta$. The partonic spin transfer coefficient becomes larger in the forward direction as well, but its contribution to the $\eta$ dependence of the asymmetry is small compared to the shape of the transversity distribution.

\begin{figure}[h]
\begin{center}
	\includegraphics[width=0.45\textwidth]{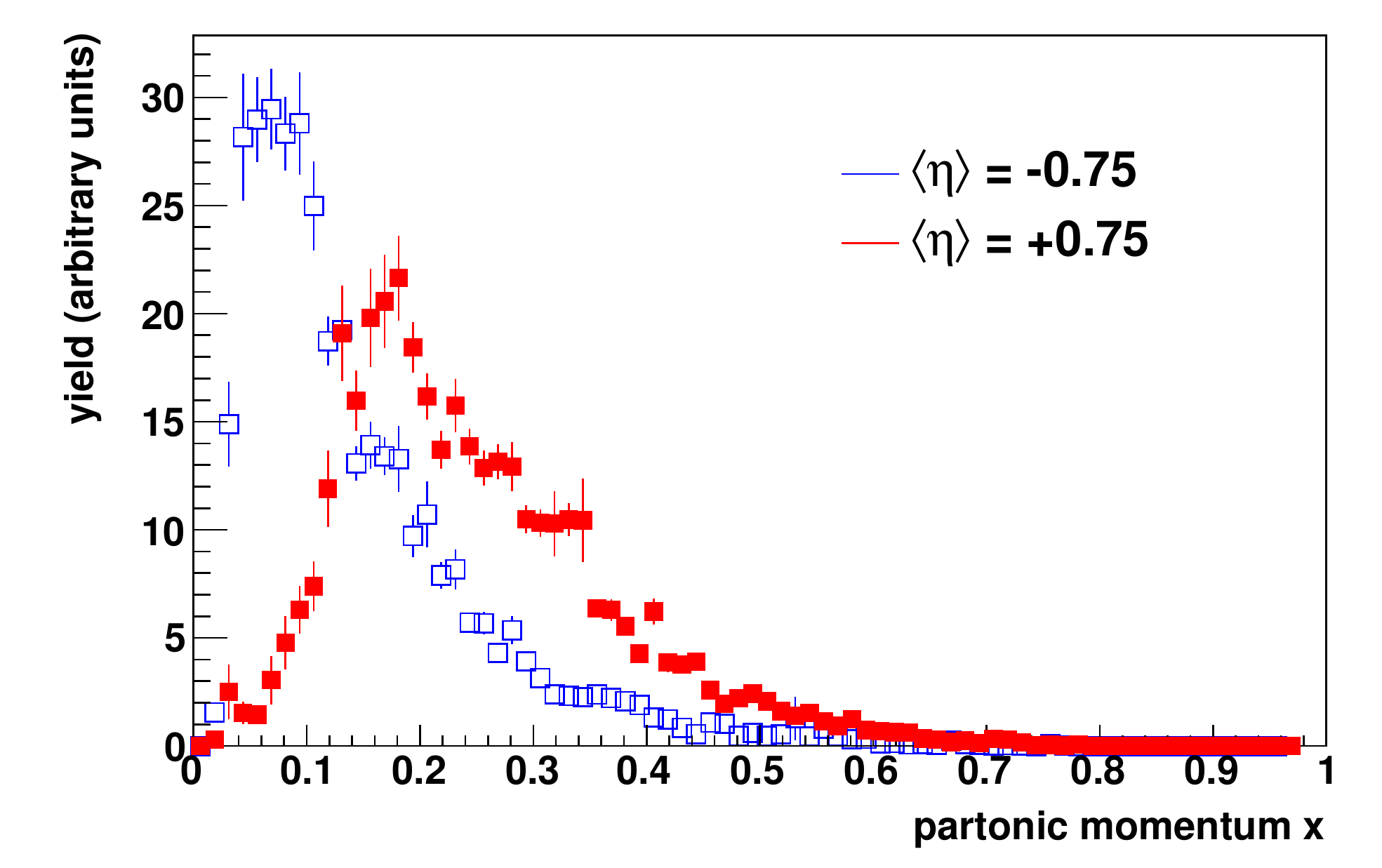}
	\caption{Comparison of the shape of the partonic momentum distributions in the polarized protons for different pseudorapidity regions of the pion pair as determined from embedded event simulation studies.
    The distributions are not fully normalized.
    \label{fig:xdist}}
\end{center}
\end{figure}

We show the corresponding distributions of $x$ in Fig.~\ref{fig:xdist} for the highest and lowest $\eta$ ranges from Fig.~\ref{fig:etaAsym}.
The distributions are fairly wide and asymmetrical as is expected for hadronic collisions.
They also partially overlap, but the different pseudrapdidities clearly are sensitive to different partonic kinematics.

\begin{figure}[h]
\begin{center}
	\includegraphics[width=0.45\textwidth]{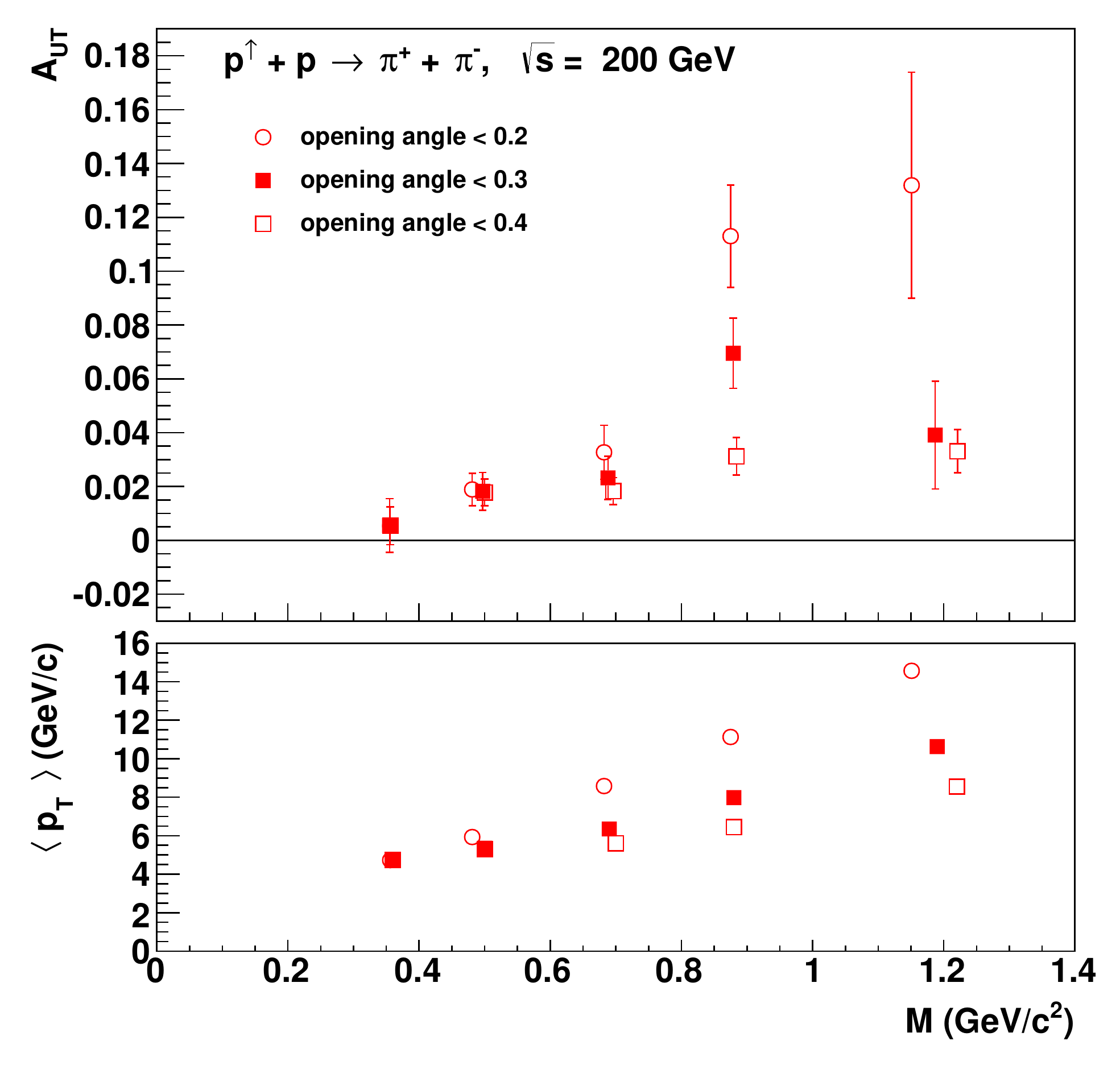}
	\caption{$A_{UT}(M)$ with different opening angle cuts. The signal in each  $M$ bin exhibits a strong dependence on the mean $p_{T}$.
Data points are slightly shifted in $M$ for better visibility.\label{fig:coneAsymM}}
\end{center}
\end{figure}

We also investigated the effect of different upper values for the opening angle, ranging from 0.2 to 0.4 as shown in Fig.~\ref{fig:coneAsymM}.
At intermediate invariant masses, $M>0.7~$GeV$/c^2$, the asymmetries depend strongly on this geometric limitation; wider opening angles result in smaller asymmetries.
As can be seen in the bottom panel, the opening angle cut also directly affects the mean $\langle p_{T}\rangle$ of the pion pair.
Larger opening angles typically allow for more pions with low momentum, so a tighter cut selects high transverse momenta; supplemental data tables list the details and are available online.
This, in effect would also increase the contributions of pion pairs with high $z$ and high $x$ which (see above) have been found to scale with the asymmetries~\cite{Vossen:2011fk}.
While there is a clear distinction between the asymmetries with increasing invariant mass, at the largest masses the statistical uncertainty, especially for the most restrictive cuts on the opening angle, does not allow for a decisive statement at this point.
It might be that the $z$-dependence of the asymmetries is a minor contribution and what we observe is actually largely driven by the change in the relative contribution of the resonant production of pions from $\rho$-decays and the non-resonant channel.
Non-resonant pions are more likely at low $p_T$ and their dominance in the data sample will dilute the asymmetry when using wider opening angle cuts.

The leading systematic uncertainty for our results comes from the ~4.8\% scale uncertainty of the beam polarization.
On average the purity of the single pion sample is 96\% which has been determined in simulation studies.
The purity shows a slight dependence on the transverse momentum, starting at around 94\% and rising up to 97\% at with $p_{T}$.
The asymmetry in $\pi-p$ correlations is expected to be very small in model calculations.
Data from $\pi-K$ asymmetries~\cite{compassPiK} are of the same sign as those of the two-pion system and of similar or smaller size.
We do not assign a systematic uncertainty to the results due to the unknown size of background asymmetry.
In the worst case the dilution of the asymmetry is on the same order of magnitude as the impurity of the pion sample. Triggering on large electromagnetic energy deposits introduces a bias in the sampled event kinematics and partonic processes~\cite{Adamczyk:2012qj}. From simulations, we determine that our trigger bias in selecting the partonic subprocess leads to an enhancement of the fraction of quark-quark scattering sampled of up to 20\% whereas quark-gluon and gluon-gluon scattering processes are suppressed by up to 10\%.
Overall, systematic uncertainties are very small compared to the statistical precision of the measurement, and they are not shown in the figures.

A variety of systematic checks have been carried out
to ensure the correctness of the results. A random assignment of the polarization states of the beam bunches leads to vanishing spin asymmetries. The $\chi^2$ values of the individual fits are distributed according to a $\chi^2$ distribution (within the relevant statistics).
An alternative way of computing the asymmetry takes advantage of the fact
that the asymmetry is antisymmetric in $\phi_{RS}$ and therefore a shift of
$\pi$ and a flip of the beam polarization both lead to a sign change of the
asymmetry~\cite{Ohlsen:1973}. The advantage of this "proper-flip" method is that the relative luminosity dependence cancels. It leads to the same result as~(\ref{eq:simpleForm}). In addition, the consistency between asymmetries for both RHIC beams is an important check, as is the stability of the results over the duration of the measurement.



In summary, STAR has observed transverse spin-dependent charged pion pair correlation asymmetries with a statistical significance of more than five standard deviations away from zero.
Using the collinear factorization framework, the distribution of transversely polarized quarks described by the proton's transversity distribution function can be extracted from these results. 
This constitutes the first signal of transversity in $p^\uparrow+p$ collisions.
The observed signal is enhanced for invariant masses of the hadron pair around the  $\rho$ mass and rises with $p_{T}$ and $\eta$ consistent with qualitative expectations from the transversity distribution function and the dependence of the IFF on $z$ and $M$.
These results can be included in an extraction of transversity from world data in a collinear framework that is currently underway~\cite{Bacchetta:2011ab}.
Compared with previous measurements of two hadron correlations in SIDIS, the RHIC data allows access to a complementary kinematic regime.
Proton-proton collisions do not suffer from $u$-quark dominance and will therefore help constrain the $d$-quark transversity. 
This is the first measurement of transverse spin asymmetries in $p^{\uparrow}+p$ collisions  which can be directly compared to those in SIDIS and $e^+-e^-$ annihilation.
Thus it is an important test of factorization and universality in these processes, particularly so in light of the recent discovery that factorization is broken for transverse spin asymmetries in $p+p$ collisions that vanish in a collinear framework~\cite{Rogers:2010dm}.

\input{acknowledgements}

\bibliography{StarIFF}
\clearpage
\onecolumngrid
\section{Supplemental Information}
\input{tables2}

\end{document}

%% file: authors.tex
\affiliation{AGH University of Science and Technology, Cracow 30-059, Poland}
\affiliation{Argonne National Laboratory, Argonne, Illinois 60439, USA}
\affiliation{Brookhaven National Laboratory, Upton, New York 11973, USA}
\affiliation{University of California, Berkeley, California 94720, USA}
\affiliation{University of California, Davis, California 95616, USA}
\affiliation{University of California, Los Angeles, California 90095, USA}
\affiliation{Central China Normal University (HZNU), Wuhan 430079, China}
\affiliation{University of Illinois at Chicago, Chicago, Illinois 60607, USA}
\affiliation{Creighton University, Omaha, Nebraska 68178, USA}
\affiliation{Czech Technical University in Prague, FNSPE, Prague, 115 19, Czech Republic}
\affiliation{Nuclear Physics Institute AS CR, 250 68 \v{R}e\v{z}/Prague, Czech Republic}
\affiliation{Frankfurt Institute for Advanced Studies FIAS, Frankfurt 60438, Germany}
\affiliation{Institute of Physics, Bhubaneswar 751005, India}
\affiliation{Indian Institute of Technology, Mumbai 400076, India}
\affiliation{Indiana University, Bloomington, Indiana 47408, USA}
\affiliation{Alikhanov Institute for Theoretical and Experimental Physics, Moscow 117218, Russia}
\affiliation{University of Jammu, Jammu 180001, India}
\affiliation{Joint Institute for Nuclear Research, Dubna, 141 980, Russia}
\affiliation{Kent State University, Kent, Ohio 44242, USA}
\affiliation{University of Kentucky, Lexington, Kentucky, 40506-0055, USA}
\affiliation{Korea Institute of Science and Technology Information, Daejeon 305-701, Korea}
\affiliation{Institute of Modern Physics, Lanzhou 730000, China}
\affiliation{Lawrence Berkeley National Laboratory, Berkeley, California 94720, USA}
\affiliation{Max-Planck-Institut fur Physik, Munich 80805, Germany}
\affiliation{Michigan State University, East Lansing, Michigan 48824, USA}
\affiliation{Moscow Engineering Physics Institute, Moscow 115409, Russia}
\affiliation{National Institute of Science Education and Research, Bhubaneswar 751005, India}
\affiliation{Ohio State University, Columbus, Ohio 43210, USA}
\affiliation{Institute of Nuclear Physics PAN, Cracow 31-342, Poland}
\affiliation{Panjab University, Chandigarh 160014, India}
\affiliation{Pennsylvania State University, University Park, Pennsylvania 16802, USA}
\affiliation{Institute of High Energy Physics, Protvino 142281, Russia}
\affiliation{Purdue University, West Lafayette, Indiana 47907, USA}
\affiliation{Pusan National University, Pusan 609735, Republic of Korea}
\affiliation{University of Rajasthan, Jaipur 302004, India}
\affiliation{Rice University, Houston, Texas 77251, USA}
\affiliation{University of Science and Technology of China, Hefei 230026, China}
\affiliation{Shandong University, Jinan, Shandong 250100, China}
\affiliation{Shanghai Institute of Applied Physics, Shanghai 201800, China}
\affiliation{Temple University, Philadelphia, Pennsylvania 19122, USA}
\affiliation{Texas A\&M University, College Station, Texas 77843, USA}
\affiliation{University of Texas, Austin, Texas 78712, USA}
\affiliation{University of Houston, Houston, Texas 77204, USA}
\affiliation{Tsinghua University, Beijing 100084, China}
\affiliation{United States Naval Academy, Annapolis, Maryland, 21402, USA}
\affiliation{Valparaiso University, Valparaiso, Indiana 46383, USA}
\affiliation{Variable Energy Cyclotron Centre, Kolkata 700064, India}
\affiliation{Warsaw University of Technology, Warsaw 00-661, Poland}
\affiliation{Wayne State University, Detroit, Michigan 48201, USA}
\affiliation{World Laboratory for Cosmology and Particle Physics (WLCAPP), Cairo 11571, Egypt}
\affiliation{Yale University, New Haven, Connecticut 06520, USA}
\affiliation{University of Zagreb, Zagreb, HR-10002, Croatia}

\author{L.~Adamczyk}\affiliation{AGH University of Science and Technology, Cracow 30-059, Poland}
\author{J.~K.~Adkins}\affiliation{University of Kentucky, Lexington, Kentucky, 40506-0055, USA}
\author{G.~Agakishiev}\affiliation{Joint Institute for Nuclear Research, Dubna, 141 980, Russia}
\author{M.~M.~Aggarwal}\affiliation{Panjab University, Chandigarh 160014, India}
\author{Z.~Ahammed}\affiliation{Variable Energy Cyclotron Centre, Kolkata 700064, India}
\author{I.~Alekseev}\affiliation{Alikhanov Institute for Theoretical and Experimental Physics, Moscow 117218, Russia}
\author{J.~Alford}\affiliation{Kent State University, Kent, Ohio 44242, USA}
\author{A.~Aparin}\affiliation{Joint Institute for Nuclear Research, Dubna, 141 980, Russia}
\author{D.~Arkhipkin}\affiliation{Brookhaven National Laboratory, Upton, New York 11973, USA}
\author{E.~C.~Aschenauer}\affiliation{Brookhaven National Laboratory, Upton, New York 11973, USA}
% ...

\collaboration{STAR Collaboration}\noaffiliation

%% file: acknowledgements.tex
We thank the RHIC Operations Group and RCF at BNL, the NERSC Center at LBNL, the KISTI Center in Korea, and the Open Science Grid consortium for providing resources and support. This work was supported in part by the Office of Nuclear Physics within the U.S. DOE Office of Science, the U.S. NSF, the Ministry of Education and Science of the Russian Federation, NNSFC, CAS, MoST and MoE of China, the Korean Research Foundation, GA and MSMT of the Czech Republic, FIAS of Germany, DAE, DST, and UGC of India, the National Science Centre of Poland, National Research Foundation, the Ministry of Science, Education and Sports of the Republic of Croatia, and RosAtom of Russia.

%% file: tables2.tex
\begin{center}
	\begin{table}[h!]
\caption{\label{trigTable1}$p_T$ asymmetries, $\eta<0$, maximum opening angle of 0.2}
		\begin{tabular}{|c|c|c|c|c|c|c|c|c|c|c|c| } \hline
\vtop{\hbox{\strut $\left<p_T\right>$}\hbox{ \strut [GeV/c] } }& $\left<\sin\theta\right>$ & \vtop{\hbox{\strut $\left<M_\textrm{Inv}\right>$}\hbox{ \strut [GeV/c$^2$] } } & $A_{\textrm{UT}}^{\sin\phi}$ & $\sigma_{A_{\textrm{UT}}^{\sin\phi}}$ & $\left<z\right>$ & $\sigma_{\left<z\right>}$ & $\left<x_1\right>$ & $\sigma_{\left<x_1\right>}$ & $\left<x_2\right>$& $\sigma_{\left<x_2\right>}$ & $\left<\eta\right>$   \\                                                                 \hline

3.62& 0.99 & 0.38 & -0.031& 0.014 & 0.29 & 0.10 &  0.15 & 0.10 & 0.20  & 0.11 & -0.50\\
4.49& 0.96 & 0.41 & -0.010& 0.012 & 0.35 & 0.12 & 0.15 & 0.10 & 0.20 & 0.11 & -0.51\\
5.70& 0.93 & 0.47 & 0.023& 0.011 &0.44 & 0.16 & 0.16 & 0.11 &  0.21 & 0.11 & -0.51\\
7.18& 0.91 & 0.55 & 0.016& 0.014 &  0.48 & 0.15 & 0.16 & 0.11& 0.23 & 0.11 & -0.51\\
10.45& 0.88 & 0.68 & 0.0085& 0.013 &  0.57 & 0.17 & 0.20 & 0.12 & 0.27 & 0.12 & -0.51\\
\hline
           			\end{tabular}
\end{table}
\end{center}

 \begin{center}
	\begin{table}[h!]
\caption{$\left<M_\textrm{Inv}\right>$ asymmetries, $\eta<0$, maximum opening angle of 0.2}
\begin{tabular}{|c|c|c|c|c|c|c|c|c|c|c|c|} \hline
 \vtop{\hbox{\strut $\left<M_\textrm{Inv}\right>$}\hbox{ \strut [GeV/c$^2$] } } & $\left<\sin\theta\right>$ & \vtop{\hbox{\strut $\left<p_T\right>$}\hbox{ \strut [GeV/c] } } & $A_{\textrm{UT}}^{\sin\phi}$ & $\sigma_{A_{\textrm{UT}}^{\sin\phi}}$ & $\left<z\right>$ & $\sigma_{\left<z\right>}$ & $\left<x_1\right>$ & $\sigma_{\left<x_1\right>}$ & $\left<x_2\right>$& $\sigma_{\left<x_2\right>}$ & $\left<\eta\right>$ \\    
 \hline                                                               
0.36& 0.95 & 4.73 & -0.013& 0.010 &  0.38 & 0.15 & 0.16 & 0.11 & 0.22 & 0.12 & -0.51\\                                                                                                   
0.48& 0.94 & 5.93 & 0.015& 0.0080 &  0.43 & 0.17 & 0.16 & 0.11 & 0.22 & 0.11 & -0.51\\                                                                                                    
0.68& 0.91 & 8.58 & -0.0044& 0.014 & 0.53 & 0.17 & 0.17 & 0.11 & 0.24 & 0.12  & -0.51\\                                                                                                   
0.88& 0.90 & 11.13 & 0.022& 0.027 &  0.59 & 0.17 & 0.21 & 0.12 & 0.28 & 0.13 & -0.51\\                                                                                                   
1.15& 0.90 & 14.57 & 0.0042& 0.062 &  0.63 & 0.15 & 0.23 & 0.13 & 0.34 & 0.13 & -0.48\\   
\hline                                                                                                		\end{tabular}
	\end{table}
\end{center}
                                                                           \begin{center}
	\begin{table}[h!]
\caption{$p_T$ asymmetries, $\eta>0$, maximum opening angle 0.2}
		\begin{tabular}{|c|c|c|c|c|c|c|c|c|c|c|c|} \hline
  \vtop{\hbox{\strut $\left<p_T\right>$}\hbox{ \strut [GeV/c] } } & $\left<\sin\theta\right>$ & \vtop{\hbox{\strut $\left<M_\textrm{Inv}\right>$}\hbox{ \strut [GeV/c$^2$] } } & $A_{\textrm{UT}}^{\sin\phi}$ & $\sigma_{A_{\textrm{UT}}^{\sin\phi}}$ & $\left<z\right>$ & $\sigma_{\left<z\right>}$ & $\left<x_1\right>$ & $\sigma_{\left<x_1\right>}$ & $\left<x_2\right>$& $\sigma_{\left<x_2\right>}$ & $\left<\eta\right>$\\
 \hline
3.62& 0.99 & 0.38 & 0.0090& 0.014 &  0.28 & 0.10 & 0.21 & 0.11 & 0.15 & 0.11 &0.50\\
4.49& 0.96 & 0.41 & 0.0095& 0.012 & 0.34 & 0.12 & 0.21 & 0.11 & 0.15 & 0.11 &0.51\\
5.70& 0.93 & 0.47 & 0.022& 0.011 & 0.42 & 0.15 & 0.22 & 0.12 & 0.15 & 0.10 &0.51\\
7.18& 0.91 & 0.55 & 0.0050& 0.014 &  0.50 & 0.19 & 0.23 & 0.12 & 0.16 & 0.10 &0.51\\
10.45& 0.88 & 0.68 & 0.057& 0.012 & 0.58 & 0.17 & 0.27 & 0.13 & 0.20 & 0.12 & 0.51\\
\hline
           		\end{tabular}
	\end{table}
\end{center}
 \begin{center}
	\begin{table}[h!]
\caption{$\left<M_\textrm{Inv}\right>$ asymmetries, $\eta>0$, maximum opening angle 0.2}
		\begin{tabular}{|c|c|c|c|c|c|c|c|c|c|c|c|} \hline
\vtop{\hbox{\strut $\left<M_\textrm{Inv}\right>$}\hbox{ \strut [GeV/c$^2$] } }  & $\left<\sin\theta\right>$ & \vtop{\hbox{\strut $\left<p_T\right>$}\hbox{ \strut [GeV/c] } }& $A_{\textrm{UT}}^{\sin\phi}$ & $\sigma_{A_{\textrm{UT}}^{\sin\phi}}$ & $\left<z\right>$ & $\sigma_{\left<z\right>}$ & $\left<x_1\right>$ & $\sigma_{\left<x_1\right>}$ & $\left<x_2\right>$& $\sigma_{\left<x_2\right>}$ & $\left<\eta\right>$   \\                                                                   \hline
0.36& 0.95 & 4.73 & 0.0054& 0.010 &  0.38 & 0.16 & 0.23 & 0.12 & 0.16 & 0.11 & 0.51\\  
0.48& 0.94 & 5.93 & 0.019& 0.0081 &  0.42 & 0.16 & 0.22 & 0.12 & 0.16 & 0.11 & 0.51\\                                                                                                     
0.68& 0.91 & 8.58 & 0.033& 0.014 &  0.55 & 0.20 & 0.25 & 0.12 & 0.17 & 0.11 & 0.51\\                                                                                                     
0.88& 0.90 & 11.13 & 0.11& 0.027 &  0.60 & 0.17 & 0.27 & 0.13 & 0.20 & 0.12 & 0.51\\                                                                                                    
1.15& 0.90 & 14.57 & 0.13& 0.060 &  0.65 & 0.17 & 0.31 & 0.13 & 0.24 & 0.13 & 0.48\\  
\hline

           		\end{tabular}
	\end{table}
\end{center}
  \begin{center}
	\begin{table}[h!]
\caption{$p_T$ asymmetries, $\eta<0$, maximum opening angle 0.3}
		\begin{tabular}{|c|c|c|c|c|c|c|c|c|c|c|c|} \hline                                                                                                                                                                                               
 \vtop{\hbox{\strut $\left<p_T\right>$}\hbox{ \strut [GeV/c] } } & $\left<\sin\theta\right>$ & \vtop{\hbox{\strut $\left<M_\textrm{Inv}\right>$}\hbox{ \strut [GeV/c$^2$] } } & $A_{\textrm{UT}}^{\sin\phi}$ & $\sigma_{A_{\textrm{UT}}^{\sin\phi}}$ & $\left<z\right>$ & $\sigma_{\left<z\right>}$ & $\left<x_1\right>$ & $\sigma_{\left<x_1\right>}$ & $\left<x_2\right>$& $\sigma_{\left<x_2\right>}$ & $\left<\eta\right>$ \\
 \hline
3.61& 0.99 & 0.45 & -0.022& 0.010 & 0.30 & 0.13 & 0.15 & 0.10 & 0.20 & 0.11  & -0.49\\
4.49& 0.97 & 0.51 & -0.007& 0.0087 & 0.36 & 0.13 & 0.15 & 0.11 & 0.20 & 0.11  & -0.50\\
5.68& 0.94 & 0.59 & 0.018& 0.0082 &  0.43 & 0.16 & 0.16 & 0.11 & 0.22 & 0.11 & -0.51\\
7.17& 0.92 & 0.67 & 0.021& 0.011 &  0.48 & 0.15 & 0.16 & 0.11 & 0.23 & 0.11 & -0.51\\
10.32& 0.88 & 0.81 & 0.0088& 0.011 & 0.57 & 0.17 & 0.19 & 0.12 & 0.27 & 0.12 & -0.51\\
\hline
           		\end{tabular}
	\end{table}
\end{center}
 \begin{center}
	\begin{table}[h!]
\caption{$\left<M_\textrm{Inv}\right>$ asymmetries, $\eta<0$, maximum opening angle 0.3}
		\begin{tabular}{|c|c|c|c|c|c|c|c|c|c|c|c|} \hline
 \vtop{\hbox{\strut $\left<M_\textrm{Inv}\right>$}\hbox{ \strut [GeV/c$^2$] } } & $\left<\sin\theta\right>$ & \vtop{\hbox{\strut $\left<p_T\right>$}\hbox{ \strut [GeV/c] } } & $A_{\textrm{UT}}^{\sin\phi}$ &$\sigma_{A_{\textrm{UT}}^{\sin\phi}}$ & $\left<z\right>$ & $\sigma_{\left<z\right>}$ & $\left<x_1\right>$ & $\sigma_{\left<x_1\right>}$ & $\left<x_2\right>$& $\sigma_{\left<x_2\right>}$ & $\left<\eta\right>$ \\           
 \hline                                                                                                                                  
0.36& 0.95 & 4.73 & -0.013& 0.010 & 0.38 & 0.15 & 0.16 & 0.11 & 0.22 & 0.12 & -0.51\\
0.50& 0.95 & 5.35 & 0.0063& 0.0068 &  0.41 & 0.17  & 0.16 & 0.11 & 0.21& 0.11 & -0.50\\ 
0.69& 0.94 & 6.35 & 0.00020 & 0.0081 &  0.44 & 0.17 & 0.16 & 0.11 & 0.22 & 0.11 & -0.51\\
0.88& 0.92 & 7.98 & 0.025& 0.013 &  0.50 & 0.17 & 0.17 & 0.11 & 0.25 & 0.12 & -0.51\\
1.19& 0.90 & 10.63 & 0.017& 0.020 & 0.56 & 0.17 & 0.19 & 0.12 & 0.27 & 0.13 & -0.50\\     
\hline                                                                                                          		\end{tabular}
	\end{table}
\end{center}                                                                         
                                                                                                                                                                                  
  \begin{center}
	\begin{table}[h!]
\caption{$p_T$ asymmetries, $\eta>0$,maximum opening angle 0.3}
		\begin{tabular}{|c|c|c|c|c|c|c|c|c|c|c|c|} \hline                                                                          
 \vtop{\hbox{\strut $\left<p_T\right>$}\hbox{ \strut [GeV/c] } }& $\left<\sin\theta\right>$ & \vtop{\hbox{\strut $\left<M_\textrm{Inv}\right>$}\hbox{ \strut [GeV/c$^2$] } } & $A_{\textrm{UT}}^{\sin\phi}$ & $\sigma_{A_{\textrm{UT}}^{\sin\phi}}$ & $\left<z\right>$ & $\sigma_{\left<z\right>}$ & $\left<x_1\right>$ & $\sigma_{\left<x_1\right>}$ & $\left<x_2\right>$& $\sigma_{\left<x_2\right>}$ & $\left<\eta\right>$  \\
 \hline
3.61& 0.99 & 0.45 & 0.0092& 0.010 &   0.29 & 0.10 & 0.20 & 0.11 & 0.14 & 0.10 & 0.49\\
4.49& 0.97 & 0.51 & 0.011& 0.0086 &   0.35 & 0.12  & 0.21 & 0.11 & 0.15 & 0.11 & 0.50\\
5.68& 0.94 & 0.59 & 0.029& 0.0082 &   0.43 & 0.15 & 0.22 & 0.12 & 0.15 & 0.10 & 0.51\\
7.17& 0.92 & 0.67 & 0.0071& 0.011 &   0.50 & 0.18 & 0.23 & 0.12 & 0.16 & 01.0 & 0.51\\
10.32& 0.88 & 0.81 & 0.053& 0.011 &  0.58 & 0.18 & 0.27 & 0.13 & 0.19 & 0.12 & 0.51\\
\hline
           		\end{tabular}
	\end{table}
\end{center}

 \begin{center}
	\begin{table}[h!]
\caption{$\left<M_\textrm{Inv}\right>$ asymmetries, $\eta>0$, maximum opening angle 0.3}
		\begin{tabular}{|c|c|c|c|c|c|c|c|c|c|c|c|} \hline
\vtop{\hbox{\strut $\left<M_\textrm{Inv}\right>$}\hbox{ \strut [GeV/c$^2$] } }  & $\left<\sin\theta\right>$ & \vtop{\hbox{\strut $\left<p_T\right>$}\hbox{ \strut [GeV/c] } } & $A_{\textrm{UT}}^{\sin\phi}$ &$\sigma_{A_{\textrm{UT}}^{\sin\phi}}$ & $\left<z\right>$ & $\sigma_{\left<z\right>}$ & $\left<x_1\right>$ & $\sigma_{\left<x_1\right>}$ & $\left<x_2\right>$& $\sigma_{\left<x_2\right>}$ & $\left<\eta\right>$ \\                                                                   \hline                                                                          
0.36& 0.95 & 4.73 & 0.0054& 0.010 &   0.38 & 0.16 & 0.23 & 0.12  & 0.16 & 0.11 & 0.51\\
0.50& 0.95 & 5.35 & 0.018& 0.0068 &  0.40  & 0.16 & 0.22 & 0.12 & 0.16 & 0.11 & 0.50\\
0.69& 0.94 & 6.35 & 0.023& 0.0081 &  0.46 & 0.18 & 0.23 & 0.12 & 0.15  & 0.11 & 0.51\\
0.88& 0.92 & 7.98 & 0.070& 0.013 &  0.51 & 0.17 & 0.24 & 0.12 & 0.17 & 0.11  & 0.51\\
1.19& 0.90 & 10.63 & 0.039& 0.020 & 0.59 & 0.19 & 0.27 & 0.12 & 0.19 & 0.12 & 0.50\\ 
\hline
           		\end{tabular}
	\end{table}
\end{center}
                                                                                                         \begin{center}
	\begin{table}[h!]
\caption{$p_T$ asymmetries, $\eta<0$, maximum opening angle 0.4}
		\begin{tabular}{|c|c|c|c|c|c|c|c|c|c|c|c|} \hline                                                     
 \vtop{\hbox{\strut $\left<p_T\right>$}\hbox{ \strut [GeV/c] } }& $\left<\sin\theta\right>$ & \vtop{\hbox{\strut $\left<M_\textrm{Inv}\right>$}\hbox{ \strut [GeV/c$^2$] } } & $A_{\textrm{UT}}^{\sin\phi}$ &$\sigma_{A_{\textrm{UT}}^{\sin\phi}}$ & $\left<z\right>$ & $\sigma_{\left<z\right>}$ & $\left<x_1\right>$ & $\sigma_{\left<x_1\right>}$ & $\left<x_2\right>$& $\sigma_{\left<x_2\right>}$ & $\left<\eta\right>$  \\
 \hline
3.60& 0.99 & 0.53 & -0.016& 0.0084 &0.30 & 0.12 & 0.14 & 0.10 & 0.20 & 0.11 & -0.49\\
4.48& 0.97 & 0.60 & -0.0081& 0.0073 &0.36 & 0.13 & 0.15 & 0.11 & 0.20 & 0.11 & -0.50\\
5.67& 0.94 & 0.68 & 0.013& 0.0073 & 0.43 & 0.16 & 0.15 & 0.11 & 0.21 & 0.11 & -0.50\\
7.17& 0.92 & 0.76 & 0.016& 0.010 & 0.48 & 0.15 & 0.16 & 0.11 & 0.24 & 0.11 &-0.51\\
10.28& 0.88 & 0.89 & 0.015& 0.0099 &0.57 & 0.17 & 0.19 & 0.12 & 0.27 & 0.12 & -0.51\\
\hline
           		\end{tabular}
	\end{table}
\end{center}
                                                             \begin{center}
	\begin{table}[h!]
\caption{$\left<M_\textrm{Inv}\right>$ asymmetries, $\eta<0$, maximum opening angle 0.4}
		\begin{tabular}{|c|c|c|c|c|c|c|c|c|c|c|c|} \hline  
		 \vtop{\hbox{\strut $\left<M_\textrm{Inv}\right>$}\hbox{ \strut [GeV/c$^2$] } }  & $\left<\sin\theta\right>$ & \vtop{\hbox{\strut $\left<p_T\right>$}\hbox{ \strut [GeV/c] } }& $A_{\textrm{UT}}^{\sin\phi}$ & $\sigma_{A_{\textrm{UT}}^{\sin\phi}}$ & $\left<z\right>$ & $\sigma_{\left<z\right>}$ & $\left<x_1\right>$ & $\sigma_{\left<x_1\right>}$ & $\left<x_2\right>$& $\sigma_{\left<x_2\right>}$ & $\left<\eta\right>$  \\                        
		 \hline                                                       
0.36& 0.95 & 4.73 & -0.013& 0.010 &0.38 & 0.15 & 0.16 & 0.11 & 0.22 & 0.12 & -0.51\\                                
0.50& 0.95 & 5.31 & 0.0054& 0.0067 & 0.41 & 0.17 & 0.16 & 0.11  & 0.21 & 0.11 & -0.50\\
0.70& 0.96 & 5.60 & -0.0022& 0.0068 & 0.41 & 0.17 & 0.16 & 0.11 & 0.22 & 0.11 & -0.50\\
0.88& 0.94 & 6.45 & 0.010& 0.0094 & 0.44 & 0.16 & 0.16 & 0.11 & 0.23 & 0.12 & -0.50\\ 
1.22& 0.91 & 8.55 & 0.011& 0.012 & 0.52 & 0.18 & 0.17 & 0.11 & 0.24 & 0.12 & -0.49     \\
\hline                                                                                
           		\end{tabular}
	\end{table}
\end{center}                                                                                      
                                                                                                                                                      \begin{center}
	\begin{table}[h!]
\caption{$p_T$ asymmetries, $\eta>0$, maximum opening angle 0.4}
		\begin{tabular}{|c|c|c|c|c|c|c|c|c|c|c|c|} \hline                                                                          
 \vtop{\hbox{\strut $\left<p_T\right>$}\hbox{ \strut [GeV/c] } } & $\left<\sin\theta\right>$ &\vtop{\hbox{\strut $\left<M_\textrm{Inv}\right>$}\hbox{ \strut [GeV/c$^2$] } }  & $A_{\textrm{UT}}^{\sin\phi}$ & $\sigma_{A_{\textrm{UT}}^{\sin\phi}}$ & $\left<z\right>$ & $\sigma_{\left<z\right>}$ & $\left<x_1\right>$ & $\sigma_{\left<x_1\right>}$ & $\left<x_2\right>$& $\sigma_{\left<x_2\right>}$ & $\left<\eta\right>$ \\
 \hline
3.60& 0.99 & 0.53 & 0.013& 0.0084 & 0.29 & 0.10 & 0.20 & 0.11 & 0.14 & 0.10 &  0.49\\
4.48& 0.97 & 0.60 & 0.0060& 0.0073& 0.35 & 0.12 & 0.21 & 0.11 & 0.15 & 0.11 &  0.50\\
5.67& 0.94 & 0.68 & 0.025& 0.0073 &0.42 & 0.14 & 0.22 & 0.12 & 0.15 & 0.10 &  0.50\\
7.17& 0.92 & 0.76 & 0.0096& 0.010 &0.50 & 0.18 & 0.23 & 0.12 & 0.16 & 0.10 &  0.51\\
10.28& 0.88 & 0.89 & 0.053& 0.0099&0.58 & 0.18 & 0.27 & 0.13 & 0.19 & 0.12 & 0.51\\
\hline
           		\end{tabular}
	\end{table}
\end{center}

 \begin{center}
	\begin{table}[h!]
\caption{$\left<M_\textrm{Inv}\right>$ asymmetries, $\eta>0$, maximum opening angle 0.4}
		\begin{tabular}{|c|c|c|c|c|c|c|c|c|c|c|c|} \hline
 \vtop{\hbox{\strut $\left<M_\textrm{Inv}\right>$}\hbox{ \strut [GeV/c$^2$] } }  & $\left<\sin\theta\right>$ & \vtop{\hbox{\strut $\left<p_T\right>$}\hbox{ \strut [GeV/c] } } & $A_{\textrm{UT}}^{\sin\phi}$ & $\sigma_{A_{\textrm{UT}}^{\sin\phi}}$ & $\left<z\right>$ & $\sigma_{\left<z\right>}$ & $\left<x_1\right>$ & $\sigma_{\left<x_1\right>}$ & $\left<x_2\right>$& $\sigma_{\left<x_2\right>}$ & $\left<\eta\right>$ \\   
 \hline                                                                                                                                          
0.36& 0.95 & 4.73 & 0.0054& 0.010 &0.38 & 0.16 & 0.23 & 0.12 & 0.16 & 0.11   & 0.51\\ 
0.50& 0.95 & 5.31 & 0.018& 0.0067 &0.39 & 0.16 & 0.22 & 0.12  & 0.16 & 0.11  & 0.50\\ 
0.70& 0.96 & 5.60 & 0.018& 0.0068 & 0.42 & 0.18 & 0.22 & 0.12 & 0.15 & 0.11  & 0.50\\ 
0.88& 0.94 & 6.45 & 0.031& 0.0095 &0.44 & 0.16 & 0.22 & 0.12 & 0.16 & 0.10   & 0.50\\ 
1.22& 0.91 & 8.55 & 0.033& 0.012 &0.53 & 0.18 & 0.24 & 0.12 & 0.17 & 0.11   & 0.49\\                                                                                          \hline
           		\end{tabular}
	\end{table}
\end{center}

 \begin{center}
	\begin{table}[h!]
\caption{$\eta$ asymmetries, maximum opening angle 0.2}
		\begin{tabular}{|c|c|c|c|c|c|c|c|c|c|c|c|} \hline
 \vtop{\hbox{\strut $\left<p_T\right>$}\hbox{ \strut [GeV/c] } }& $\left<\sin\theta\right>$ & \vtop{\hbox{\strut $\left<M_\textrm{Inv}\right>$}\hbox{ \strut [GeV/c$^2$] } }  & $A_{\textrm{UT}}^{\sin\phi}$ & $\sigma_{A_{\textrm{UT}}^{\sin\phi}}$ & $\left<z\right>$ & $\sigma_{\left<z\right>}$ & $\left<x_1\right>$ & $\sigma_{\left<x_1\right>}$ & $\left<x_2\right>$& $\sigma_{\left<x_2\right>}$ & $\left<\eta\right>$ \\     
 \hline
 6.30 & 0.93 & 0.50 & 0.0010 &0.0079 &   0.46 & 0.18 & 0.15 & 0.11 & 0.25 & 0.12 & -0.75 \\               
 6.27 & 0.94 & 0.50 & 0.0054 &0.0080 &   0.41 & 0.16 & 0.18 & 0.11 & 0.20 & 0.11 & -0.26 \\
 6.27 & 0.94 & 0.50 & 0.015   &0.0080 &   0.43 & 0.18 & 0.21 & 0.12 & 0.17 & 0.11 & 0.26\\
 6.30 & 0.93 & 0.50 & 0.028   &0.0079 &   0.45 & 0.19 & 0.26 & 0.12 & 0.15 & 0.11 & 0.75  \\
 \hline
                       		\end{tabular}
	\end{table}
\end{center}           
       
        \begin{center}
	\begin{table}[h!]
\caption{$\eta$ asymmetries, maximum opening angle 0.3}
		\begin{tabular}{|c|c|c|c|c|c|c|c|c|c|c|c|} \hline
 \vtop{\hbox{\strut $\left<p_T\right>$}\hbox{ \strut [GeV/c] } }& $\left<\sin\theta\right>$ & \vtop{\hbox{\strut $\left<M_\textrm{Inv}\right>$}\hbox{ \strut [GeV/c$^2$] } }  & $A_{\textrm{UT}}^{\sin\phi}$ & $\sigma_{A_{\textrm{UT}}^{\sin\phi}}$ & $\left<z\right>$ & $\sigma_{\left<z\right>}$ & $\left<x_1\right>$ & $\sigma_{\left<x_1\right>}$ & $\left<x_2\right>$& $\sigma_{\left<x_2\right>}$ & $\left<\eta\right>$\\  
\hline
 6.04 & 0.94 & 0.59 & -0.00020 &0.0060&0.45 & 0.18 & 0.15 & 0.11 & 0.24 & 0.12  & -0.75\\ 
 5.99 & 0.94 & 0.60 & 0.0075 &0.0061& 0.40 & 0.16 & 0.18 & 0.11 & 0.20 & 0.11 & -0.26 \\
 5.99 & 0.94 & 0.60 & 0.013 &0.0061& 0.42 & 0.17 & 0.20 & 0.11 & 0.17 & 0.11 & 0.26  \\
 6.04 & 0.94 & 0.59 & 0.031 &0.0060& 0.44 & 0.18 & 0.25 & 0.12 & 0.15 & 0.10  &0.75 \\
 \hline
 \end{tabular}
	\end{table}
\end{center}           
 \begin{center}
	\begin{table}[h!]
\caption{\label{trigTableN}$\eta$ asymmetries, maximum opening angle 0.4}
		\begin{tabular}{|c|c|c|c|c|c|c|c|c|c|c|c|} \hline
		\vtop{\hbox{\strut $\left<p_T\right>$}\hbox{ \strut [GeV/c] } }& $\left<\sin\theta\right>$ & \vtop{\hbox{\strut $\left<M_\textrm{Inv}\right>$}\hbox{ \strut [GeV/c$^2$] } }  &$A_{\textrm{UT}}^{\sin\phi}$ & $\sigma_{A_{\textrm{UT}}^{\sin\phi}}$ & $\left<z\right>$ & $\sigma_{\left<z\right>}$ & $\left<x_1\right>$ & $\sigma_{\left<X_1\right>}$ & $\left<x_2\right>$& $\sigma_{\left<x_2\right>}$ & $\left<\eta\right>$\\
		\hline
5.84 & 0.95 & 0.67 & -0.0027 &0.0053&0.44 & 0.18 & 0.14 & 0.11 & 0.24 & 0.12 &-0.75    \\
5.78 & 0.95 & 0.67 & 0.0074 &0.0053& 0.40 & 0.16 & 0.17 & 0.11 & 0.20 & 0.11 &-0.26 \\
5.78 & 0.95 & 0.67 & 0.0095 &0.0053& 0.41 & 0.17 & 0.20 & 0.11 & 0.17 & 0.11 &0.26 \\
5.84 & 0.95 & 0.67 & 0.030 &0.0053& 0.43 & 0.18 & 0.24 & 0.12 & 0.14 & 0.10 & 0.75 \\
\hline
\end{tabular}
	\end{table}
\end{center}

%% file: main.bbl
\begin{thebibliography}{33}%
\makeatletter
\providecommand \@ifxundefined [1]{%
 \@ifx{#1\undefined}
}%
\providecommand \@ifnum [1]{%
 \ifnum #1\expandafter \@firstoftwo
 \else \expandafter \@secondoftwo
 \fi
}%
\providecommand \@ifx [1]{%
 \ifx #1\expandafter \@firstoftwo
 \else \expandafter \@secondoftwo
 \fi
}%
\providecommand \natexlab [1]{#1}%
\providecommand \enquote  [1]{``#1''}%
\providecommand \bibnamefont  [1]{#1}%
\providecommand \bibfnamefont [1]{#1}%
\providecommand \citenamefont [1]{#1}%
\providecommand \href@noop [0]{\@secondoftwo}%
\providecommand \href [0]{\begingroup \@sanitize@url \@href}%
\providecommand \@href[1]{\@@startlink{#1}\@@href}%
\providecommand \@@href[1]{\endgroup#1\@@endlink}%
\providecommand \@sanitize@url [0]{\catcode `\\12\catcode `\$12\catcode
  `\&12\catcode `\#12\catcode `\^12\catcode `\_12\catcode `\%12\relax}%
\providecommand \@@startlink[1]{}%
\providecommand \@@endlink[0]{}%
\providecommand \url  [0]{\begingroup\@sanitize@url \@url }%
\providecommand \@url [1]{\endgroup\@href {#1}{\urlprefix }}%
\providecommand \urlprefix  [0]{URL }%
\providecommand \Eprint [0]{\href }%
\providecommand \doibase [0]{http://dx.doi.org/}%
\providecommand \selectlanguage [0]{\@gobble}%
\providecommand \bibinfo  [0]{\@secondoftwo}%
\providecommand \bibfield  [0]{\@secondoftwo}%
\providecommand \translation [1]{[#1]}%
\providecommand \BibitemOpen [0]{}%
\providecommand \bibitemStop [0]{}%
\providecommand \bibitemNoStop [0]{.\EOS\space}%
\providecommand \EOS [0]{\spacefactor3000\relax}%
\providecommand \BibitemShut  [1]{\csname bibitem#1\endcsname}%
\let\auto@bib@innerbib\@empty
\bibitem [{\citenamefont {Barone}\ \emph {et~al.}(2002)\citenamefont {Barone},
  \citenamefont {Drago},\ and\ \citenamefont {Ratcliffe}}]{Barone:2001sp}%
  \BibitemOpen
  \bibfield  {author} {\bibinfo {author} {\bibfnamefont {V.}~\bibnamefont
  {Barone}}, \bibinfo {author} {\bibfnamefont {A.}~\bibnamefont {Drago}}, \
  and\ \bibinfo {author} {\bibfnamefont {P.~G.}\ \bibnamefont {Ratcliffe}},\
  }\href {\doibase 10.1016/S0370-1573(01)00051-5} {\bibfield  {journal}
  {\bibinfo  {journal} {Phys.Rept.}\ }\textbf {\bibinfo {volume} {359}},\
  \bibinfo {pages} {1} (\bibinfo {year} {2002})},\ \Eprint
  {http://arxiv.org/abs/hep-ph/0104283} {arXiv:hep-ph/0104283 [hep-ph]}
  \BibitemShut {NoStop}%
\bibitem [{\citenamefont {Anselmino}\ \emph {et~al.}(2013)\citenamefont
  {Anselmino}, \citenamefont {Boglione}, \citenamefont {D'Alesio},
  \citenamefont {Melis}, \citenamefont {Murgia} \emph
  {et~al.}}]{Anselmino:2013vqa}%
  \BibitemOpen
  \bibfield  {author} {\bibinfo {author} {\bibfnamefont {M.}~\bibnamefont
  {Anselmino}}, \bibinfo {author} {\bibfnamefont {M.}~\bibnamefont {Boglione}},
  \bibinfo {author} {\bibfnamefont {U.}~\bibnamefont {D'Alesio}}, \bibinfo
  {author} {\bibfnamefont {S.}~\bibnamefont {Melis}}, \bibinfo {author}
  {\bibfnamefont {F.}~\bibnamefont {Murgia}},  \emph {et~al.},\ }\href
  {\doibase 10.1103/PhysRevD.87.094019} {\bibfield  {journal} {\bibinfo
  {journal} {Phys.Rev.}\ }\textbf {\bibinfo {volume} {D87}},\ \bibinfo {pages}
  {094019} (\bibinfo {year} {2013})},\ \Eprint {http://arxiv.org/abs/1303.3822}
  {arXiv:1303.3822 [hep-ph]} \BibitemShut {NoStop}%
\bibitem [{\citenamefont {Gamberg}\ and\ \citenamefont
  {Goldstein}(2001)}]{Gamberg:2001qc}%
  \BibitemOpen
  \bibfield  {author} {\bibinfo {author} {\bibfnamefont {L.~P.}\ \bibnamefont
  {Gamberg}}\ and\ \bibinfo {author} {\bibfnamefont {G.~R.}\ \bibnamefont
  {Goldstein}},\ }\href {\doibase 10.1103/PhysRevLett.87.242001} {\bibfield
  {journal} {\bibinfo  {journal} {Phys.Rev.Lett.}\ }\textbf {\bibinfo {volume}
  {87}},\ \bibinfo {pages} {242001} (\bibinfo {year} {2001})},\ \Eprint
  {http://arxiv.org/abs/hep-ph/0107176} {arXiv:hep-ph/0107176 [hep-ph]}
  \BibitemShut {NoStop}%
\bibitem [{\citenamefont {Bacchetta}\ \emph {et~al.}(2013)\citenamefont
  {Bacchetta}, \citenamefont {Courtoy},\ and\ \citenamefont
  {Radici}}]{Bacchetta:2012ty}%
  \BibitemOpen
  \bibfield  {author} {\bibinfo {author} {\bibfnamefont {A.}~\bibnamefont
  {Bacchetta}}, \bibinfo {author} {\bibfnamefont {A.}~\bibnamefont {Courtoy}},
  \ and\ \bibinfo {author} {\bibfnamefont {M.}~\bibnamefont {Radici}},\ }\href
  {\doibase 10.1007/JHEP03(2013)119} {\bibfield  {journal} {\bibinfo  {journal}
  {JHEP}\ }\textbf {\bibinfo {volume} {1303}},\ \bibinfo {pages} {119}
  (\bibinfo {year} {2013})},\ \Eprint {http://arxiv.org/abs/1212.3568}
  {arXiv:1212.3568} \BibitemShut {NoStop}%
\bibitem [{\citenamefont {Cloet}\ \emph {et~al.}(2008)\citenamefont {Cloet},
  \citenamefont {Bentz},\ and\ \citenamefont {Thomas}}]{Cloet:2007em}%
  \BibitemOpen
  \bibfield  {author} {\bibinfo {author} {\bibfnamefont {I.}~\bibnamefont
  {Cloet}}, \bibinfo {author} {\bibfnamefont {W.}~\bibnamefont {Bentz}}, \ and\
  \bibinfo {author} {\bibfnamefont {A.~W.}\ \bibnamefont {Thomas}},\ }\href
  {\doibase 10.1016/j.physletb.2007.09.071} {\bibfield  {journal} {\bibinfo
  {journal} {Phys.Lett.}\ }\textbf {\bibinfo {volume} {B659}},\ \bibinfo
  {pages} {214} (\bibinfo {year} {2008})},\ \Eprint
  {http://arxiv.org/abs/0708.3246} {arXiv:0708.3246 [hep-ph]} \BibitemShut
  {NoStop}%
\bibitem [{\citenamefont {Wakamatsu}(2007)}]{Wakamatsu:2007nc}%
  \BibitemOpen
  \bibfield  {author} {\bibinfo {author} {\bibfnamefont {M.}~\bibnamefont
  {Wakamatsu}},\ }\href {\doibase 10.1016/j.physletb.2007.08.013} {\bibfield
  {journal} {\bibinfo  {journal} {Phys.Lett.}\ }\textbf {\bibinfo {volume}
  {B653}},\ \bibinfo {pages} {398} (\bibinfo {year} {2007})},\ \Eprint
  {http://arxiv.org/abs/0705.2917} {arXiv:0705.2917 [hep-ph]} \BibitemShut
  {NoStop}%
\bibitem [{\citenamefont {Gockeler}\ \emph {et~al.}(2005)\citenamefont
  {Gockeler} \emph {et~al.}}]{Gockeler:2005cj}%
  \BibitemOpen
  \bibfield  {author} {\bibinfo {author} {\bibfnamefont {M.}~\bibnamefont
  {Gockeler}} \emph {et~al.} (\bibinfo {collaboration} {QCDSF Collaboration,
  UKQCD Collaboration}),\ }\href {\doibase 10.1016/j.physletb.2005.09.002}
  {\bibfield  {journal} {\bibinfo  {journal} {Phys.Lett.}\ }\textbf {\bibinfo
  {volume} {B627}},\ \bibinfo {pages} {113} (\bibinfo {year} {2005})},\ \Eprint
  {http://arxiv.org/abs/hep-lat/0507001} {arXiv:hep-lat/0507001 [hep-lat]}
  \BibitemShut {NoStop}%
\bibitem [{\citenamefont {He}\ and\ \citenamefont {Ji}(1996)}]{He:1996wy}%
  \BibitemOpen
  \bibfield  {author} {\bibinfo {author} {\bibfnamefont {H.-x.}\ \bibnamefont
  {He}}\ and\ \bibinfo {author} {\bibfnamefont {X.-D.}\ \bibnamefont {Ji}},\
  }\href {\doibase 10.1103/PhysRevD.54.6897} {\bibfield  {journal} {\bibinfo
  {journal} {Phys.Rev.}\ }\textbf {\bibinfo {volume} {D54}},\ \bibinfo {pages}
  {6897} (\bibinfo {year} {1996})},\ \Eprint
  {http://arxiv.org/abs/hep-ph/9607408} {arXiv:hep-ph/9607408 [hep-ph]}
  \BibitemShut {NoStop}%
\bibitem [{\citenamefont {Schweitzer}\ \emph {et~al.}(2001)\citenamefont
  {Schweitzer}, \citenamefont {Urbano}, \citenamefont {Polyakov}, \citenamefont
  {Weiss}, \citenamefont {Pobylitsa} \emph {et~al.}}]{Schweitzer:2001sr}%
  \BibitemOpen
  \bibfield  {author} {\bibinfo {author} {\bibfnamefont {P.}~\bibnamefont
  {Schweitzer}}, \bibinfo {author} {\bibfnamefont {D.}~\bibnamefont {Urbano}},
  \bibinfo {author} {\bibfnamefont {M.~V.}\ \bibnamefont {Polyakov}}, \bibinfo
  {author} {\bibfnamefont {C.}~\bibnamefont {Weiss}}, \bibinfo {author}
  {\bibfnamefont {P.}~\bibnamefont {Pobylitsa}},  \emph {et~al.},\ }\href
  {\doibase 10.1103/PhysRevD.64.034013} {\bibfield  {journal} {\bibinfo
  {journal} {Phys.Rev.}\ }\textbf {\bibinfo {volume} {D64}},\ \bibinfo {pages}
  {034013} (\bibinfo {year} {2001})},\ \Eprint
  {http://arxiv.org/abs/hep-ph/0101300} {arXiv:hep-ph/0101300 [hep-ph]}
  \BibitemShut {NoStop}%
\bibitem [{\citenamefont {Pasquini}\ \emph {et~al.}(2005)\citenamefont
  {Pasquini}, \citenamefont {Pincetti},\ and\ \citenamefont
  {Boffi}}]{Pasquini:2005dk}%
  \BibitemOpen
  \bibfield  {author} {\bibinfo {author} {\bibfnamefont {B.}~\bibnamefont
  {Pasquini}}, \bibinfo {author} {\bibfnamefont {M.}~\bibnamefont {Pincetti}},
  \ and\ \bibinfo {author} {\bibfnamefont {S.}~\bibnamefont {Boffi}},\ }\href
  {\doibase 10.1103/PhysRevD.72.094029} {\bibfield  {journal} {\bibinfo
  {journal} {Phys.Rev.}\ }\textbf {\bibinfo {volume} {D72}},\ \bibinfo {pages}
  {094029} (\bibinfo {year} {2005})},\ \Eprint
  {http://arxiv.org/abs/hep-ph/0510376} {arXiv:hep-ph/0510376 [hep-ph]}
  \BibitemShut {NoStop}%
\bibitem [{\citenamefont {Bacchetta}\ \emph {et~al.}(2008)\citenamefont
  {Bacchetta}, \citenamefont {Conti},\ and\ \citenamefont
  {Radici}}]{Bacchetta:2008af}%
  \BibitemOpen
  \bibfield  {author} {\bibinfo {author} {\bibfnamefont {A.}~\bibnamefont
  {Bacchetta}}, \bibinfo {author} {\bibfnamefont {F.}~\bibnamefont {Conti}}, \
  and\ \bibinfo {author} {\bibfnamefont {M.}~\bibnamefont {Radici}},\ }\href
  {\doibase 10.1103/PhysRevD.78.074010} {\bibfield  {journal} {\bibinfo
  {journal} {Phys.Rev.}\ }\textbf {\bibinfo {volume} {D78}},\ \bibinfo {pages}
  {074010} (\bibinfo {year} {2008})},\ \Eprint {http://arxiv.org/abs/0807.0323}
  {arXiv:0807.0323 [hep-ph]} \BibitemShut {NoStop}%
\bibitem [{\citenamefont {Airapetian}\ \emph {et~al.}(2005)\citenamefont
  {Airapetian} \emph {et~al.}}]{Airapetian:2004tw}%
  \BibitemOpen
  \bibfield  {author} {\bibinfo {author} {\bibfnamefont {A.}~\bibnamefont
  {Airapetian}} \emph {et~al.} (\bibinfo {collaboration} {HERMES
  Collaboration}),\ }\href {\doibase 10.1103/PhysRevLett.94.012002} {\bibfield
  {journal} {\bibinfo  {journal} {Phys.Rev.Lett.}\ }\textbf {\bibinfo {volume}
  {94}},\ \bibinfo {pages} {012002} (\bibinfo {year} {2005})},\ \Eprint
  {http://arxiv.org/abs/hep-ex/0408013} {arXiv:hep-ex/0408013 [hep-ex]}
  \BibitemShut {NoStop}%
\bibitem [{\citenamefont {Airapetian}\ \emph {et~al.}(2008)\citenamefont
  {Airapetian} \emph {et~al.}}]{Airapetian:2008sk}%
  \BibitemOpen
  \bibfield  {author} {\bibinfo {author} {\bibfnamefont {A.}~\bibnamefont
  {Airapetian}} \emph {et~al.} (\bibinfo {collaboration} {HERMES
  Collaboration}),\ }\href {\doibase 10.1088/1126-6708/2008/06/017} {\bibfield
  {journal} {\bibinfo  {journal} {JHEP}\ }\textbf {\bibinfo {volume} {0806}},\
  \bibinfo {pages} {017} (\bibinfo {year} {2008})},\ \Eprint
  {http://arxiv.org/abs/0803.2367} {arXiv:0803.2367 [hep-ex]} \BibitemShut
  {NoStop}%
\bibitem [{\citenamefont {Adolph}\ \emph {et~al.}(2012)\citenamefont {Adolph}
  \emph {et~al.}}]{Adolph:2012nw}%
  \BibitemOpen
  \bibfield  {author} {\bibinfo {author} {\bibfnamefont {C.}~\bibnamefont
  {Adolph}} \emph {et~al.} (\bibinfo {collaboration} {COMPASS Collaboration}),\
  }\href {\doibase 10.1016/j.physletb.2012.05.015} {\bibfield  {journal}
  {\bibinfo  {journal} {Phys.Lett.}\ }\textbf {\bibinfo {volume} {B713}},\
  \bibinfo {pages} {10} (\bibinfo {year} {2012})},\ \Eprint
  {http://arxiv.org/abs/1202.6150} {arXiv:1202.6150 [hep-ex]} \BibitemShut
  {NoStop}%
\bibitem [{\citenamefont {Alekseev}\ \emph {et~al.}(2010)\citenamefont
  {Alekseev} \emph {et~al.}}]{Alekseev:2010rw}%
  \BibitemOpen
  \bibfield  {author} {\bibinfo {author} {\bibfnamefont {M.}~\bibnamefont
  {Alekseev}} \emph {et~al.} (\bibinfo {collaboration} {COMPASS
  Collaboration}),\ }\href {\doibase 10.1016/j.physletb.2010.08.001} {\bibfield
   {journal} {\bibinfo  {journal} {Phys.Lett.}\ }\textbf {\bibinfo {volume}
  {B692}},\ \bibinfo {pages} {240} (\bibinfo {year} {2010})},\ \Eprint
  {http://arxiv.org/abs/1005.5609} {arXiv:1005.5609 [hep-ex]} \BibitemShut
  {NoStop}%
\bibitem [{\citenamefont {Alekseev}\ \emph {et~al.}(2009)\citenamefont
  {Alekseev} \emph {et~al.}}]{Alekseev:2008aa}%
  \BibitemOpen
  \bibfield  {author} {\bibinfo {author} {\bibfnamefont {M.}~\bibnamefont
  {Alekseev}} \emph {et~al.} (\bibinfo {collaboration} {COMPASS
  Collaboration}),\ }\href {\doibase 10.1016/j.physletb.2009.01.060} {\bibfield
   {journal} {\bibinfo  {journal} {Phys.Lett.}\ }\textbf {\bibinfo {volume}
  {B673}},\ \bibinfo {pages} {127} (\bibinfo {year} {2009})},\ \Eprint
  {http://arxiv.org/abs/0802.2160} {arXiv:0802.2160 [hep-ex]} \BibitemShut
  {NoStop}%
\bibitem [{\citenamefont {Seidl}\ \emph {et~al.}(2008)\citenamefont {Seidl}
  \emph {et~al.}}]{Seidl:2008xc}%
  \BibitemOpen
  \bibfield  {author} {\bibinfo {author} {\bibfnamefont {R.}~\bibnamefont
  {Seidl}} \emph {et~al.} (\bibinfo {collaboration} {Belle Collaboration}),\
  }\href {\doibase 10.1103/PhysRevD.78.032011, 10.1103/PhysRevD.86.039905}
  {\bibfield  {journal} {\bibinfo  {journal} {Phys.Rev.}\ }\textbf {\bibinfo
  {volume} {D78}},\ \bibinfo {pages} {032011} (\bibinfo {year} {2008})},\
  \Eprint {http://arxiv.org/abs/0805.2975} {arXiv:0805.2975 [hep-ex]}
  \BibitemShut {NoStop}%
\bibitem [{\citenamefont {Vossen}\ \emph {et~al.}(2011)\citenamefont {Vossen}
  \emph {et~al.}}]{Vossen:2011fk}%
  \BibitemOpen
  \bibfield  {author} {\bibinfo {author} {\bibfnamefont {A.}~\bibnamefont
  {Vossen}} \emph {et~al.} (\bibinfo {collaboration} {Belle Collaboration}),\
  }\href {\doibase 10.1103/PhysRevLett.107.072004} {\bibfield  {journal}
  {\bibinfo  {journal} {Phys.Rev.Lett.}\ }\textbf {\bibinfo {volume} {107}},\
  \bibinfo {pages} {072004} (\bibinfo {year} {2011})},\ \Eprint
  {http://arxiv.org/abs/1104.2425} {arXiv:1104.2425 [hep-ex]} \BibitemShut
  {NoStop}%
\bibitem [{\citenamefont {Rogers}\ and\ \citenamefont
  {Mulders}(2010)}]{Rogers:2010dm}%
  \BibitemOpen
  \bibfield  {author} {\bibinfo {author} {\bibfnamefont {T.~C.}\ \bibnamefont
  {Rogers}}\ and\ \bibinfo {author} {\bibfnamefont {P.~J.}\ \bibnamefont
  {Mulders}},\ }\href {\doibase 10.1103/PhysRevD.81.094006} {\bibfield
  {journal} {\bibinfo  {journal} {Phys.Rev.}\ }\textbf {\bibinfo {volume}
  {D81}},\ \bibinfo {pages} {094006} (\bibinfo {year} {2010})},\ \Eprint
  {http://arxiv.org/abs/1001.2977} {arXiv:1001.2977 [hep-ph]} \BibitemShut
  {NoStop}%
\bibitem [{\citenamefont {Jaffe}\ \emph {et~al.}(1998)\citenamefont {Jaffe},
  \citenamefont {Jin},\ and\ \citenamefont {Tang}}]{PhysRevLett.80.1166}%
  \BibitemOpen
  \bibfield  {author} {\bibinfo {author} {\bibfnamefont {R.~L.}\ \bibnamefont
  {Jaffe}}, \bibinfo {author} {\bibfnamefont {X.}~\bibnamefont {Jin}}, \ and\
  \bibinfo {author} {\bibfnamefont {J.}~\bibnamefont {Tang}},\ }\href {\doibase
  10.1103/PhysRevLett.80.1166} {\bibfield  {journal} {\bibinfo  {journal}
  {Phys. Rev. Lett.}\ }\textbf {\bibinfo {volume} {80}},\ \bibinfo {pages}
  {1166} (\bibinfo {year} {1998})}\BibitemShut {NoStop}%
\bibitem [{\citenamefont {Boer}\ \emph {et~al.}(2003)\citenamefont {Boer},
  \citenamefont {Jakob},\ and\ \citenamefont {Radici}}]{Boer:2003ya}%
  \BibitemOpen
  \bibfield  {author} {\bibinfo {author} {\bibfnamefont {D.}~\bibnamefont
  {Boer}}, \bibinfo {author} {\bibfnamefont {R.}~\bibnamefont {Jakob}}, \ and\
  \bibinfo {author} {\bibfnamefont {M.}~\bibnamefont {Radici}},\ }\href
  {\doibase 10.1103/PhysRevD.67.094003} {\bibfield  {journal} {\bibinfo
  {journal} {Phys.Rev.}\ }\textbf {\bibinfo {volume} {D67}},\ \bibinfo {pages}
  {094003} (\bibinfo {year} {2003})},\ \Eprint
  {http://arxiv.org/abs/hep-ph/0302232} {arXiv:hep-ph/0302232 [hep-ph]}
  \BibitemShut {NoStop}%
\bibitem [{\citenamefont {{RHIC polarimeter group}}(2013)}]{polarimeter}%
  \BibitemOpen
  \bibfield  {author} {\bibinfo {author} {\bibnamefont {{RHIC polarimeter
  group}}},\ }\href@noop {} {\bibfield  {journal} {\bibinfo  {journal}
  {RHIC/CAD Physics Note}\ }\textbf {\bibinfo {volume} {490}} (\bibinfo {year}
  {2013})}\BibitemShut {NoStop}%
\bibitem [{\citenamefont {Ackermann}\ \emph {et~al.}(2003)\citenamefont
  {Ackermann} \emph {et~al.}}]{Ackermann:2002ad}%
  \BibitemOpen
  \bibfield  {author} {\bibinfo {author} {\bibfnamefont {K.}~\bibnamefont
  {Ackermann}} \emph {et~al.} (\bibinfo {collaboration} {STAR Collaboration}),\
  }\href {\doibase 10.1016/S0168-9002(02)01960-5} {\bibfield  {journal}
  {\bibinfo  {journal} {Nucl.Instrum.Meth.}\ }\textbf {\bibinfo {volume}
  {A499}},\ \bibinfo {pages} {624} (\bibinfo {year} {2003})}\BibitemShut
  {NoStop}%
\bibitem [{\citenamefont {Beddo}\ \emph {et~al.}(2003)\citenamefont {Beddo}
  \emph {et~al.}}]{Beddo:2002zx}%
  \BibitemOpen
  \bibfield  {author} {\bibinfo {author} {\bibfnamefont {M.}~\bibnamefont
  {Beddo}} \emph {et~al.} (\bibinfo {collaboration} {STAR Collaboration}),\
  }\href {\doibase 10.1016/S0168-9002(02)01970-8} {\bibfield  {journal}
  {\bibinfo  {journal} {Nucl.Instrum.Meth.}\ }\textbf {\bibinfo {volume}
  {A499}},\ \bibinfo {pages} {725} (\bibinfo {year} {2003})}\BibitemShut
  {NoStop}%
\bibitem [{\citenamefont {Da}\ \emph {et~al.}(2011)\citenamefont {Da} \emph
  {et~al.}}]{Da:2011da}%
  \BibitemOpen
  \bibfield  {author} {\bibinfo {author} {\bibfnamefont {H.}~\bibnamefont {Da}}
  \emph {et~al.},\ }\href@noop {} {\  (\bibinfo {year} {2011})},\ \Eprint
  {http://arxiv.org/abs/1112.2946} {arXiv:1112.2946 [hep-ex]} \BibitemShut
  {NoStop}%
\bibitem [{\citenamefont {Bacchetta}\ and\ \citenamefont
  {Radici}(2004)}]{Bacchetta:2004it}%
  \BibitemOpen
  \bibfield  {author} {\bibinfo {author} {\bibfnamefont {A.}~\bibnamefont
  {Bacchetta}}\ and\ \bibinfo {author} {\bibfnamefont {M.}~\bibnamefont
  {Radici}},\ }\href {\doibase 10.1103/PhysRevD.70.094032} {\bibfield
  {journal} {\bibinfo  {journal} {Phys.Rev.}\ }\textbf {\bibinfo {volume}
  {D70}},\ \bibinfo {pages} {094032} (\bibinfo {year} {2004})},\ \Eprint
  {http://arxiv.org/abs/hep-ph/0409174} {arXiv:hep-ph/0409174 [hep-ph]}
  \BibitemShut {NoStop}%
\bibitem [{\citenamefont {Sjostrand}\ \emph {et~al.}(2006)\citenamefont
  {Sjostrand}, \citenamefont {Mrenna},\ and\ \citenamefont
  {Skands}}]{Sjostrand:2006za}%
  \BibitemOpen
  \bibfield  {author} {\bibinfo {author} {\bibfnamefont {T.}~\bibnamefont
  {Sjostrand}}, \bibinfo {author} {\bibfnamefont {S.}~\bibnamefont {Mrenna}}, \
  and\ \bibinfo {author} {\bibfnamefont {P.~Z.}\ \bibnamefont {Skands}},\
  }\href {\doibase 10.1088/1126-6708/2006/05/026} {\bibfield  {journal}
  {\bibinfo  {journal} {JHEP}\ }\textbf {\bibinfo {volume} {0605}},\ \bibinfo
  {pages} {026} (\bibinfo {year} {2006})},\ \Eprint
  {http://arxiv.org/abs/hep-ph/0603175} {arXiv:hep-ph/0603175 [hep-ph]}
  \BibitemShut {NoStop}%
\bibitem [{\citenamefont {Brun}\ \emph {et~al.}(1978)\citenamefont {Brun},
  \citenamefont {Hagelberg}, \citenamefont {Hansroul},\ and\ \citenamefont
  {Lassalle}}]{Brun:1978fy}%
  \BibitemOpen
  \bibfield  {author} {\bibinfo {author} {\bibfnamefont {R.}~\bibnamefont
  {Brun}}, \bibinfo {author} {\bibfnamefont {R.}~\bibnamefont {Hagelberg}},
  \bibinfo {author} {\bibfnamefont {M.}~\bibnamefont {Hansroul}}, \ and\
  \bibinfo {author} {\bibfnamefont {J.}~\bibnamefont {Lassalle}},\ }\href@noop
  {} {\  (\bibinfo {year} {1978})}\BibitemShut {NoStop}%
\bibitem [{\citenamefont {Bianconi}\ \emph {et~al.}(2000)\citenamefont
  {Bianconi}, \citenamefont {Boffi}, \citenamefont {Jakob},\ and\ \citenamefont
  {Radici}}]{Bianconi:1999uc}%
  \BibitemOpen
  \bibfield  {author} {\bibinfo {author} {\bibfnamefont {A.}~\bibnamefont
  {Bianconi}}, \bibinfo {author} {\bibfnamefont {S.}~\bibnamefont {Boffi}},
  \bibinfo {author} {\bibfnamefont {R.}~\bibnamefont {Jakob}}, \ and\ \bibinfo
  {author} {\bibfnamefont {M.}~\bibnamefont {Radici}},\ }\href {\doibase
  10.1103/PhysRevD.62.034009} {\bibfield  {journal} {\bibinfo  {journal}
  {Phys.Rev.}\ }\textbf {\bibinfo {volume} {D62}},\ \bibinfo {pages} {034009}
  (\bibinfo {year} {2000})},\ \Eprint {http://arxiv.org/abs/hep-ph/9907488}
  {arXiv:hep-ph/9907488 [hep-ph]} \BibitemShut {NoStop}%
\bibitem [{\citenamefont {Braun}(2013)}]{compassPiK}%
  \BibitemOpen
  \bibfield  {author} {\bibinfo {author} {\bibfnamefont {C.}~\bibnamefont
  {Braun}}\ }(\bibinfo  {publisher} {{presented at the workshop Structure of
  Nucleons and Nuclei, Como, 10-14 June}},\ \bibinfo {year} {2013})\BibitemShut
  {NoStop}%
\bibitem [{\citenamefont {Adamczyk}\ \emph {et~al.}(2012)\citenamefont
  {Adamczyk} \emph {et~al.}}]{Adamczyk:2012qj}%
  \BibitemOpen
  \bibfield  {author} {\bibinfo {author} {\bibfnamefont {L.}~\bibnamefont
  {Adamczyk}} \emph {et~al.} (\bibinfo {collaboration} {STAR Collaboration}),\
  }\href {\doibase 10.1103/PhysRevD.86.032006} {\bibfield  {journal} {\bibinfo
  {journal} {Phys.Rev.}\ }\textbf {\bibinfo {volume} {D86}},\ \bibinfo {pages}
  {032006} (\bibinfo {year} {2012})},\ \Eprint {http://arxiv.org/abs/1205.2735}
  {arXiv:1205.2735 [nucl-ex]} \BibitemShut {NoStop}%
\bibitem [{\citenamefont {Ohlsen}\ and\ \citenamefont
  {Keaton}(1973)}]{Ohlsen:1973}%
  \BibitemOpen
  \bibfield  {author} {\bibinfo {author} {\bibfnamefont {G.~G.}\ \bibnamefont
  {Ohlsen}}\ and\ \bibinfo {author} {\bibfnamefont {P.~W.}\ \bibnamefont
  {Keaton}},\ }\href@noop {} {\bibfield  {journal} {\bibinfo  {journal} {Nucl.
  Instrum. Meth.}\ }\textbf {\bibinfo {volume} {109}} (\bibinfo {year}
  {1973})}\BibitemShut {NoStop}%
\bibitem [{\citenamefont {Bacchetta}\ \emph {et~al.}(2011)\citenamefont
  {Bacchetta}, \citenamefont {Courtoy},\ and\ \citenamefont
  {Radici}}]{Bacchetta:2011ab}%
  \BibitemOpen
  \bibfield  {author} {\bibinfo {author} {\bibfnamefont {A.}~\bibnamefont
  {Bacchetta}}, \bibinfo {author} {\bibfnamefont {A.}~\bibnamefont {Courtoy}},
  \ and\ \bibinfo {author} {\bibfnamefont {M.}~\bibnamefont {Radici}},\ }\href
  {\doibase 10.1103/PhysRevLett.107.012001} {\bibfield  {journal} {\bibinfo
  {journal} {Phys. Rev. Lett.}\ }\textbf {\bibinfo {volume} {107}},\ \bibinfo
  {pages} {012001} (\bibinfo {year} {2011})}\BibitemShut {NoStop}%
\end{thebibliography}%
